\documentclass[journal=jacsat,manuscript=communication,etalmode=truncate,keywords=true]{achemso}
\setkeys{acs}{articletitle=true,maxauthors=10,etalmode=truncate,keywords=true}
\usepackage{amsmath,amssymb,units,txfonts}
\usepackage{dblfloatfix}
\usepackage{amsfonts}

\newcommand{\Ticus}{Ti$_{\textrm{cus}}$}
\usepackage{graphicx}
\usepackage{helvet,wasysym}
\usepackage[usenames,dvipsnames]{color}
\definecolor{StyleColor}{RGB}{11,122,64} % JCTC
\definecolor{LinkColor}{RGB}{34,80,169} % JPCC & JCTC
\definecolor{abstractcolor}{RGB}{255,243,201} % JPCC & JCTC

\usepackage[bookmarks=true,colorlinks=true,urlcolor=LinkColor,linkcolor=LinkColor,citecolor=LinkColor]{hyperref} 
\usepackage{colortbl,cuted,bm}
\usepackage{multirow}
\usepackage{xr}
\newcommand{\Hmat}{\mathcal{H}}

\newcommand{\V}{\mathcal{V}}
\newcommand{\W}{\mathcal{W}}
\newcommand{\myint}{{\int}}
\usepackage[utf8x]{inputenc}
\usepackage{xfrac}
\makeatletter\newenvironment{abstractbox}{%
   \begin{lrbox}{\@tempboxa}\begin{minipage}{0.988\textwidth}}{\end{minipage}\end{lrbox}%
   \colorbox{abstractcolor}{\usebox{\@tempboxa}}
}\makeatother

\renewcommand*\authorfont{\flushleft}%\bf}
\renewcommand*\affilfont{\mdseries\flushleft\textrm}
\renewcommand*\titlesize{\Large}
\renewcommand*\titlefont{\flushleft\sf\bfseries}
\def\refname{\sffamily\color{StyleColor}REFERENCES}
\usepackage{titlesec}
\titleformat{\section}{\bfseries\sffamily\color{StyleColor}}{\thesection.~}{0pt}{}
\titleformat{\subsection}[runin]{\bfseries\sffamily\normalsize}{\indent\thesubsection.~}{0pt}{}[.]
\titlespacing{\subsection}{0pt}{0pt}{*1}
\titleformat{\subsubsection}{\bfseries\sffamily\normalsize}{\thethesubsection.~}{0pt}{}
\titlespacing{\subsubsection}{0pt}{0pt}{*0}

\title{Optical Absorption Spectra and Excitons of Dye-Substrate Interfaces: Catechol on TiO$_{\text{2}}$(110)}
\author{Duncan John Mowbray}
\email{duncan.mowbray@gmail.com}
\affiliation[UPV/EHU]{\newline\footnotemark[1]{\ } Nano-Bio Spectroscopy Group and ETSF Scientific Development Center, Departamento de F{\'{\i}}sica de Materiales, Universidad del Pa{\'{\i}}s Vasco UPV/EHU and DIPC, E-20018 San Sebasti\'{a}n, Spain}
\author{Annapaola Migani}
\email{annapaola.migani@icn2.es}
\affiliation[ICN2CSIC]{\newline\footnotemark[3]{\ } Catalan Institute of Nanoscience and Nanotechnology (ICN2), CSIC and The Barcelona Institute of Science and Technology, Campus UAB, Bellaterra, E-08193 Barcelona, Spain}

\begin{document}
%\setpagewiselinenumbers
%\linenumbers
%\maketitle

\begin{strip}
\vspace{-1.cm}

\noindent{\color{StyleColor}{\rule{\textwidth}{0.5pt}}}
\begin{abstractbox}
\begin{tabular*}{17cm}{b{11.5cm}r}
%% Abstract Format:
%% 1. What is the problem?
%% 2. Why is it interesting?
%% 3. How do we tackle the problem?
%% 4. How is this different from previous methods?
%% 5. Does our theoretical method reproduce experimental results?
%% 6. What important results to we get?
%% 7. How do our results impact the community as a whole?
\noindent\textbf{\color{StyleColor}{ABSTRACT:}}
Optimizing the photovoltaic efficiency of dye-sensitized solar cells (DSSC) based on staggered gap heterojunctions requires a detailed understanding of sub-band gap transitions in the visible from the dye directly to the substrate's conduction band (CB) (type-II DSSCs).  Here, we calculate the optical absorption spectra and spatial distribution of bright excitons in the visible region for a prototypical DSSC, catechol on rutile TiO$_2$(110), as a function of coverage and deprotonation of the OH anchoring groups.  This is accomplished by solving the Bethe-Salpeter equation (BSE) based on hybrid range-separated exchange and correlation functional (HSE06) density functional theory (DFT) calculations.  Such a treatment is necessary to accurately describe the interfacial level alignment and the weakly bound charge transfer transitions that are the dominant absorption mechanism in type-II DSSCs.  Our HSE06 BSE spectra agree semi-quantitatively with spectra measured for catechol on anatase TiO$_2$ nanoparticles.  Our results suggest deprotonation of catechol's OH anchoring groups, while being nearly isoenergetic at high coverages, shifts the onset of 
&\includegraphics[width=5.1cm]{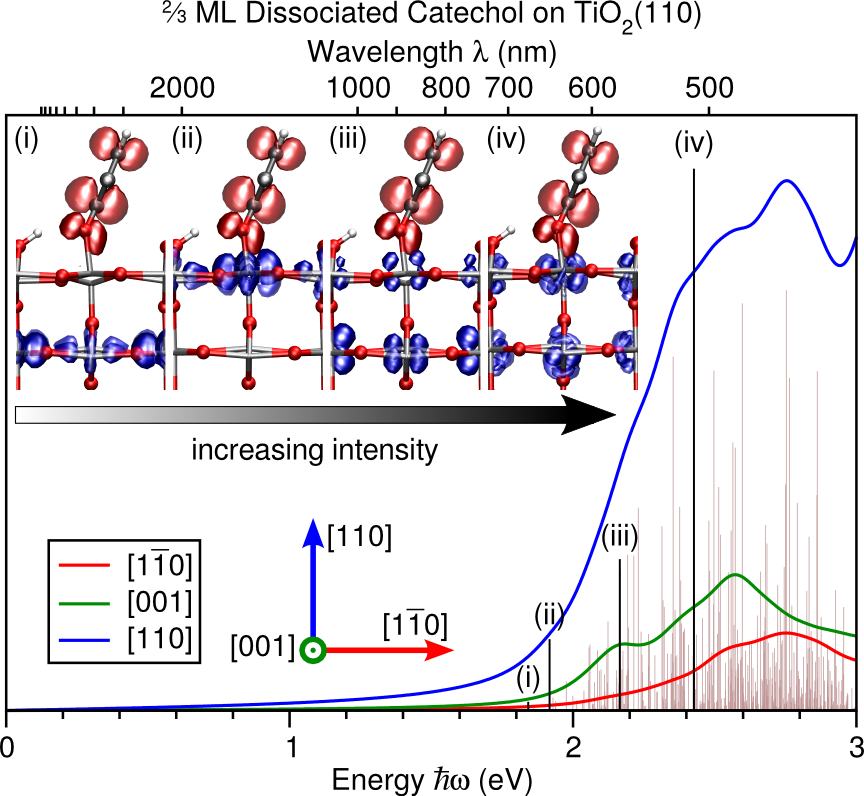}
\\
\multicolumn{2}{p{17cm}}{
the absorption spectra to lower energies, with a concomitant increase in photovoltaic efficiency.  Further, the most relevant bright excitons in the visible region are rather intense charge transfer transitions with the electron and hole spatially separated in both the [110] and [001] directions.  Such detailed information on the absorption spectra and excitons is only accessible via periodic models of the combined dye-substrate interface.
%
%\noindent\textbf{\color{StyleColor}{Keywords:}} BSE $\cdot$ HSE06 $\cdot$ DSSC 
}
\end{tabular*}
\end{abstractbox}
\noindent{\color{StyleColor}{\rule{\textwidth}{0.5pt}}}
\end{strip}

\def\bigfirstletter#1#2{{\noindent
    \setbox0\hbox{{\color{StyleColor}{\Huge #1}}}\setbox1\hbox{#2}\setbox2\hbox{(}%
    \count0=\ht0\advance\count0 by\dp0\count1\baselineskip
    \advance\count0 by-\ht1\advance\count0 by\ht2
    \dimen1=.5ex\advance\count0 by\dimen1\divide\count0 by\count1
    \advance\count0 by1\dimen0\wd0
    \advance\dimen0 by.25em\dimen1=\ht0\advance\dimen1 by-\ht1
    \global\hangindent\dimen0\global\hangafter-\count0
    \hskip-\dimen0\setbox0\hbox to\dimen0{\raise-\dimen1\box0\hss}%
    \dp0=0in\ht0=0in\box0}#2}

\section{INTRODUCTION}

Dye-sensitized solar cells (DSSCs) show great promise as sources of inexpensive, flexible, clean energy \cite{GratzelNature2001}.  The photovoltaic efficiency of such a device is proportional to its absorption of the solar spectrum\cite{PhotovoltaicEfficiency}.  The introduction of the dye facilitates absorption within the visible region via sub-band gap transitions \cite{MaromJPCLett2014,DeAngelisDSSC}.  For DSSCs with staggered gap heterojunctions, this is via transitions from the highest occupied molecular orbital (HOMO) of the dye in the substrate's gap to the conduction band (CB) \cite{CalzolariJACS2011}.  In such DSSCs, an electron is excited either to the lowest unoccupied molecular orbital (LUMO) of the dye and relaxes to the conduction band minimum (CBM) of the substrate (type-I DSSC)\cite{ModelingDyeCellsACR2012,ModelingDyeCellsJPCL2013,DyeTiO2InterfaceACSMaterials2013,SelloniChemRev2014,DeAngelisDSSCTypeIJPCC2015}, or to the CBM directly (type-II DSSC)\cite{CTTypeIvsTypeII,TypeIIvsI,CatecholTiO2ReviewChemPhysChem2012,CatecholTiO2NPJPCB2000,CatecholTiO2NPReviewEES2014,SelloniCatecholTDDFT}.  For type-I DSSCs, the device's photovoltaic efficiency is determined by the dye's energy gap and the interfacial level alignment of the LUMO and the CBM\cite{DeAngelisDSSCTypeIJPCC2015}, while for type-II DSSCs, the device's photovoltaic efficiency is determined by the interfacial level alignment of the HOMO and the CBM\cite{CatecholTiO2NPJPCB2000}.  To determine which process dominates, one must consider the optical absorption of the isolated dye, substrate, and their interface.

A prototypical staggered gap heterojunction is the catechol--TiO$_2$ interface, \cite{GratzelExpCatecholTiO2NP,CatecholTiO2NPTJPCB2003,CatecholTiO2NPJPCC2012,CatecholTiO2NPReviewEES2014,SelloniCatecholTDDFT,CatecholTiO2NPJPCB2000,Diebold,RisplendiPhysChemChemPhys2013,PrezdhoAnnRevPhysChem2006,MaromJPCLett2014,Catechol,DieboldCatecholScience,PrezhdoJPhysChemB2005,SelloniJACS2011,CTexcitonsCatecholAnatase,CatecholTiO2ReviewChemPhysChem2012} as catechol is a typical anchoring group for more complex organic and inorganic dyes \cite{CatecholAnchoringGroup,TypeIIvsI,CalzolariJACS2011,CatecholTiO2ReviewChemPhysChem2012}, and TiO$_2$ is a highly stable large gap semiconductor commonly used in staggered gap heterojunctions.  Furthermore, for catechol-TiO$_2$ interfaces, other factors that reduce the device's photovoltaic efficiency, such as electron-hole recombination, binding and conductivity, are less relevant.  For this reason, the photovoltaic efficiency of catechol--TiO$_2$ interfaces is mostly determined by the strength of the optical absorption in the visible region.

The level alignment of catechol overlayers on rutile TiO$_2$(110) has been studied experimentally via ultraviolet photoemission spectroscopy (UPS) \cite{Rangan20104829,Diebold}, inverse photoemission spectroscopy (IPES) \cite{Rangan20104829}, and theoretically using density functional theory (DFT) \cite{Diebold,RisplendiPhysChemChemPhys2013,PrezdhoAnnRevPhysChem2006,MaromJPCLett2014,Catechol,Batista,SelloniCatecholTDDFT} and quasiparticle (QP) techniques such as $G_0W_0$ \cite{MaromJPCLett2014,Catechol}.  We have previously shown\cite{OurJACS,MiganiLong,MiganiInvited,MiganiH2OJCTC2014,SunH2OAnatase,Catechol} that $G_0W_0$ \cite{GW,AngelGWReview,KresseG0W0} calculations based on the PBE\cite{PBE} exchange and correlation (xc)-functional reproduce the interfacial level alignment of molecular levels from UPS\cite{Rangan20104829,Diebold,ThorntonH2ODissTiO2110,DeSegovia,Krischok,UPSMethanol1998}, IPES\cite{Rangan20104829}, and two-photon photoemission (2PP)\cite{Onda200532} experiments.  In so doing, we were able to verify the structure of catechol overlayers on the TiO$_2$(110) surface\cite{Catechol}. 

However, to describe the efficiency of a photovoltaic device, one also needs to consider the intensity of absorption of the solar spectrum.  Here, we calculate the optical absorption spectra and excitons of  catechol--TiO$_2$(110) interfaces by solving the Bethe-Salpeter equation (BSE)\cite{BetheSalpeterEqn,KresseBSE} based on hybrid range-separated (HSE06)\cite{HSE06} xc-functional DFT calculations, using an isotropic screening of the electron-hole interaction ($\epsilon_\infty \approx 4$)
\cite{PRBKresse2009TDHSE,KresseRPA}.

Since catechol's intramolecular HOMO$\rightarrow$LUMO transitions on TiO$_2$ are supra-band gap by more than 2 eV\cite{Catechol}, the only sub-band gap transitions are from the HOMO and HOMO$-1$ levels of the catechol overlayer\cite{Diebold,Catechol} to the unoccupied Ti 3$d$ levels with $t_{2g}$-like symmetry\cite{DuncanTiO2,Selloni2PP,PetekPRB2015} of the substrate. These are charge transfer transitions\cite{CTexcitonsCatecholAnatase,GratzelExpCatecholTiO2NP,CatecholTiO2NPJPCC2012,CatecholTiO2NPTJPCB2003,CatecholTiO2NPReviewEES2014}, which require a proper description of the long-ranged electron-hole interaction\cite{LluisCT,ZwijenburgJCTC2014a,ZwijenburgJCTC2014b,NunziJCTC2015},  as provided by hybrid xc-functionals \cite{CTHSE,KresseRPA,PRBKresse2009TDHSE}. It is the level alignment of these levels, and the distribution of their wave functions, which is most relevant for describing the optical absorption of the catechol--TiO$_2$(110) interface at the BSE level.  

Our previous work has shown that $G_0W_0$ calculations significantly overestimate the electronic band gap of the TiO$_2$(110) surface by $\sim 1$~eV \cite{MiganiLong}.  This makes the quasiparticle $G_0W_0$ eigenvalues of the catechol--TiO$_2$(110) interface a poor starting point for solving the Bethe-Salpeter equation.  Conversely, the electronic band gap of the TiO$_2$(110) surface is described near quantitatively by DFT calculations based on the hybrid range-separated HSE06 xc-functional\cite{MiganiLong}.  Furthermore, HSE06 DFT calculations provide Kohn-Sham (KS) wave functions in closer agreement to the many-body wave functions obtained via self-consistent quasiparticle $GW$ (scQP$GW$) calculations\cite{OurJACS,MiganiLong}.  This suggests HSE06 DFT may provide a better basis set for BSE calculations of absorption spectra.  This is important, as we are restricting our electron-hole basis set to occupied HOMO and HOMO$-1$ levels and unoccupied levels less than 7 eV above in our BSE calculations.  

We begin by providing the computational parameters employed in section~\ref{Subsect:ComputationalDetails}, the terminology used in section~\ref{Subsect:Terminology}, the procedure for aligning the density of states (DOS) in section~\ref{Subsect:DOSAlignment}, and the Bethe-Salpeter equation in section~\ref{Subsect:BSE}.  For the most stable catechol overlayers on rutile TiO$_2$(110) we provide their structure and stability in section~\ref{Subsect:Adsorption}, interfacial level alignment in section~\ref{Subsect:Alignment}, and sub-band gap optical absorption in section~\ref{Subsect:Absorption}.  This is followed by concluding remarks in section~\ref{Sect:Conclusions}.  

\section{METHODOLOGY}\label{Sect:Methodology}

\subsection{Computational Details}\label{Subsect:ComputationalDetails}
All DFT calculations have been performed using the \textsc{vasp} code within the projector augmented wave (PAW) scheme \cite{kresse1999}, employing the HSE06\cite{HSE06} range-separated hybrid exchange and correlation (xc)-functional.   The geometries were taken from ref.~\citenum{Catechol}, where they were fully relaxed using the PBE\cite{PBE} xc-functional, with all forces $\lesssim$ 0.02 eV/\AA.  To consider the effect of long-ranged van der Waals dispersion forces on the overlayer's structure and adsorption energy, we employ the vdW-DF\cite{vdW-DF,vdW-DF2,vdW-DFVASP} xc-functional based on PBE exchange.   We employ a plane-wave energy cutoff of 445 eV, an electronic temperature $k_B T\approx0.2$ eV with all energies extrapolated to $T\rightarrow 0$ K, that is, excluding the electronic entropy contribution to the free energy $-ST$, and a PAW pseudopotential for Ti that includes the 3$s^2$ and 3$p^6$ semi-core levels.  The calculations have been performed spin unpolarized.  

All unit cells contain a four layer TiO$_2$(110) slab, employ the measured lattice parameters of bulk rutile TiO$_2$  ($a=4.5941$ \AA, $c=2.958$ \AA)\cite{TiO2LatticeParameters}, and include at least 27 \AA\ of vacuum between repeated images ($L_z=50$ \AA{} in [110] direction). For 1 and \sfrac{2}{3} monolayer (ML) catechol coverages, equivalent  $1\times4$ and $1\times3$ catechol overlayers are adsorbed on both sides of the slab, yielding a total of four and two catechol molecules, respectively.  We employ $\Gamma$ centered 4$\times$3$\times$1 $k$-point meshes, with densities $\Delta k < 0.25$~\AA$^{-1}$, and approximately 9\sfrac{1}{4} unoccupied bands per atom, that is, including all levels up to 30~eV above the valence band maximum (VBM).  

For comparison, PBE $G_0W_0$ quasiparticle calculations have been performed based on the KS eigenvalues and wave functions from PBE DFT, as previously described in ref~\citenum{Catechol}.  The $G_0W_0$ parameters employed are consistent with those successfully used to describe both bulk rutile and anatase TiO$_2$, rutile TiO$_2$(110) and anatase TiO$_2$(101) clean surfaces, and their interfaces\cite{OurJACS,MiganiLong,SunH2OAnatase,MiganiH2OJCTC2014,Catechol,MiganiInvited}.

We solve the Bethe-Salpeter equation using the KS eigenvalues and wave functions from HSE06 DFT as input, and a constant frequency-independent dielectric function ($\varepsilon_\infty \approx 4$). For the BSE calculations we have used 480 frequency sampling points, and included all the transitions from occupied HOMO and HOMO$-1$ levels to unoccupied levels with a maximum energy difference of 7 eV.\cite{KresseBSE}  In so doing, the calculated spectra include absorption by the adsorbed dye, and transfer to conduction levels of the substrate.  

For interfaces, the calculated intensities are inversely proportional to the unit cell parameter normal to the surface, that is, $I_\lambda \propto 1/L_z$.   Since we employ the same unit cell parameter along the [110] direction for each catechol--TiO$_2$(110) structure, $L_z = 50$~\AA, although the units of the computed intensities are necessarily arbitrary, they are the same for each of the four catechol--TiO$_2$(110) interfaces considered herein.

\subsection{Terminology}\label{Subsect:Terminology}

Throughout the manuscript, the atomic orbitals of Ti are described in the octahedral basis, with $x$ and $y$ axes aligned with the Ti--O bonds, that is, according to  $O_h$ crystal field theory\cite{Selloni2PP}.  In this way, the atomic orbitals are grouped into those with $t_{2g}$-like symmetry, $d_{xy}$, $d_{xz}$, and $d_{y z}$, and those with $e_g$-like symmetry, $d_{x^2-y^2}$ and $d_{z^2}$.

Following the definitions of refs.~\citenum{Catechol} and \citenum{CatecholCorrection}, the adsorption energy per molecule of catechol $E_{\textrm{ads}}$ on coordinately unsaturated Ti (\Ticus{}) sites of a TiO$_2$(110) surface is given by
\begin{align}
E_{\textrm{ads}} \approx& \frac{E[N_{\textrm{cat}}\textrm{C}_6\textrm{H}_4(\textrm{OH})_2+\textrm{TiO}_2\textrm{(110)}] - E[\textrm{TiO}_2\textrm{(110)}]}{N_{\textrm{cat}}}\nonumber\\
&- E[\textrm{C}_6\textrm{H}_4(\textrm{OH})_2].
\end{align}
Here $N_{\textrm{cat}}$ is the number of adsorbed catechol molecules in the unit cell,
and $E[N_{\textrm{cat}}\textrm{C}_6\textrm{H}_4(\textrm{OH})_2+\textrm{TiO}_2\textrm{(110)}]$,
$E[\textrm{TiO}_2\textrm{(110)}]$, and $E[\textrm{C}_6\textrm{H}_4(\textrm{OH})_2]$ are the total
energies of the covered and clean surfaces and gas phase catechol
molecule, respectively.  We model catechol in the gas phase using the most stable conformation, which has an intramolecular hydrogen bond, that is, $C_S$ symmetry. \cite{Catechol}

\subsection{Density of States Alignment}\label{Subsect:DOSAlignment}

Following the procedure laid out in ref.~\citenum{Catechol}, we plot the DOS for catechol overlayers relative to the VBM of the clean surface $\varepsilon_{\mathrm{VBM}}^{\mathrm{clean}}$.  This is accomplished by aligning the deepest Ti 3$s^2$ semi-core levels  for the clean and catechol covered TiO$_2$(110) surfaces, $\varepsilon_{\textrm{Ti} 3s^2}^{\textrm{clean}}$ and $\varepsilon_{\textrm{Ti} 3s^2}$, respectively.   In this way, $\varepsilon_{\mathrm{VBM}} = \varepsilon_{\mathrm{VBM}}^{\textrm{clean}} - \varepsilon_{\textrm{Ti} 3s^2}^{\textrm{clean}} + \varepsilon_{\textrm{Ti} 3s^2}$.  This allows us to directly compare the spectra of catechol overlayers as a function of coverage and deprotonation of the OH anchoring groups, that is, dissociation. \cite{Catechol}

\subsection{Bethe-Salpter Equation}\label{Subsect:BSE}

\begin{figure*}[!t]
%\noindent{\color{StyleColor}{\rule{\textwidth}{1pt}}}
\includegraphics[width=\textwidth]{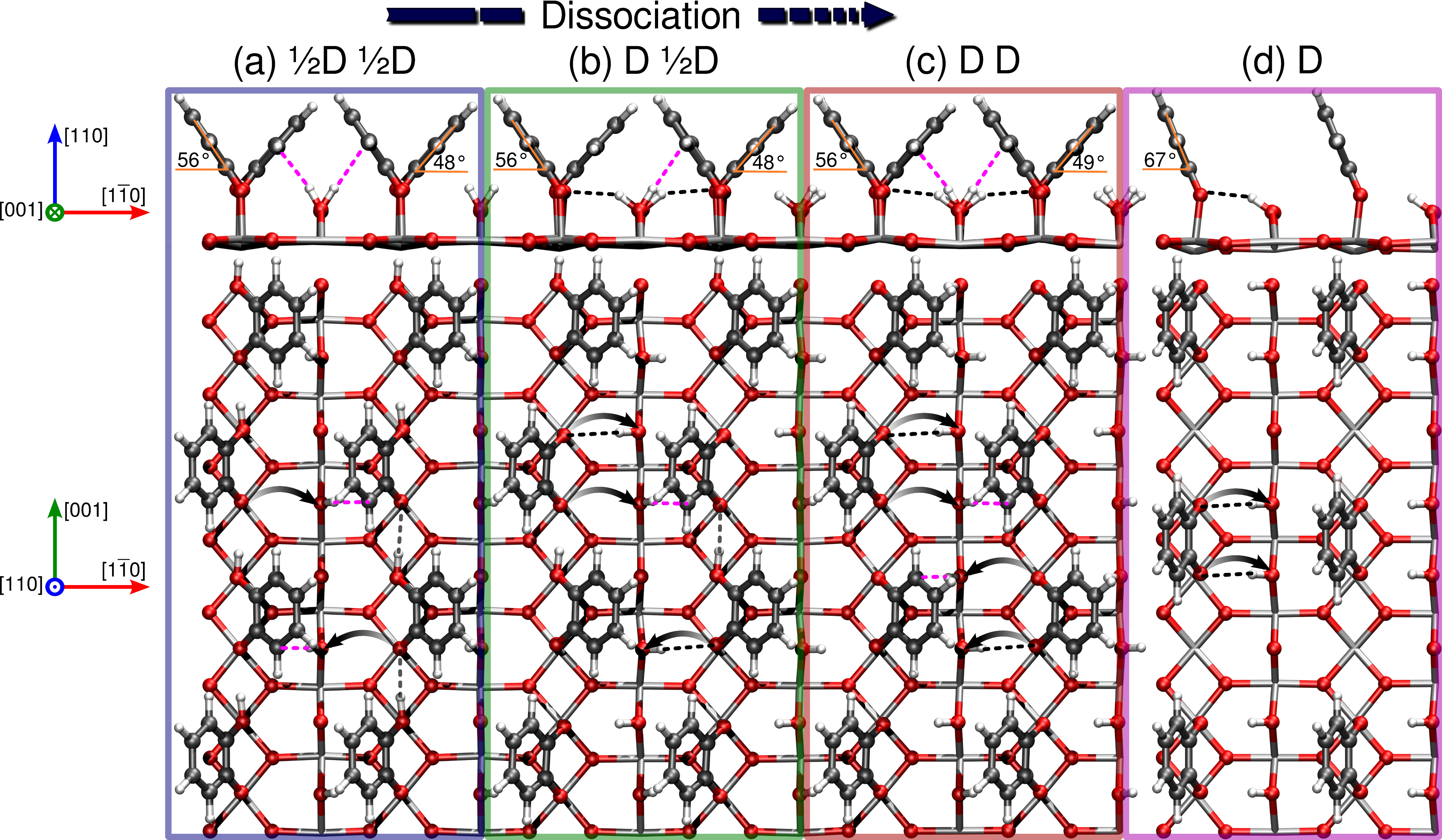}
\caption{Schematics of 1 ML catechol adsorbed (a) half (\sfrac{1}{2}D \sfrac{1}{2}D, blue), (b) mixed (D \sfrac{1}{2}D, green), (c) fully (D D, red) dissociated and (d) \sfrac{2}{3} ML (violet) catechol adsorbed dissociated (D) on \Ticus{} sites of TiO$_2$(110).  Charge transfer of $\sim-0.4 e$ accompanying deprotonation is represented by arrows, while intermolecular OH--O (gray) and interfacial O$_{\textrm{br}}$H--O (black) and O$_{\textrm{br}}$H--C (magenta) hydrogen bonds are denoted by dotted lines.  The angle between catechol's benzene ring and the surface plane, $\theta$, is shown above.  Adapted with permission from ref.~\citenum{Catechol}. Copyright 2015 American Chemical Society.
}\label{Structures:fgr}
\noindent{\color{StyleColor}{\rule{\textwidth}{1pt}}}
\end{figure*}

Assuming the KS wave functions $\psi_n$, with eigenenergies $\varepsilon_n$ and occupations $f_n$, form an orthonormal and complete basis set, then in their electron-hole pair space, suppressing their $k$-point dependence, the four-point Bethe-Salpeter equation for the density-density response function $\chi$ is\cite{BetheSalpeterEqn,AngelGWReview,KresseBSE}
\begin{equation}
\chi_{SS'} = \chi_S^0 \left[\delta_{n n'}\delta_{mm'} + \sum_{S''} \left[\V_{SS''} - \W_{SS''}\right]\chi_{S''S'}(\omega)\right]\label{BSE}
\end{equation}
where $m,m' \in \{1,\ldots,N_{\textrm{occ}}\}$, $n,n' \in \{N_{\textrm{occ}}+1,\ldots,N_{\textrm{tot}}\}$,  $N_{\textrm{occ}}$ is the number of occupied levels, $N_{\textrm{tot}}$ is the total number of electronic levels included, $S\equiv \{m,n\}$, $\chi_{nm}^0$ is the non-interacting density-density response function
\begin{equation}
\chi_{nm}^0 = \frac{f_m - f_n}{\omega - \varepsilon_n + \varepsilon_m} \delta_{n n'}\delta_{mm'}\label{chi0}
\end{equation}
$\V_{SS''}$ is the Coulomb interaction
\begin{equation}
\V_{SS''} = \myint \textrm{d}\textbf{r}\textrm{d}\textbf{r}'\frac{\psi_m(\textbf{r})\psi_{n}^*(\textbf{r}) {\psi_{m}^*}_{''}(\textbf{r}')\psi_{n''}(\textbf{r}')}{\|\textbf{r}-\textbf{r}'\|}\label{Coulomb}
\end{equation}
and $\W_{SS''}$ is the static $(\omega=0)$ screening
\begin{equation}
%%\W_{SS''} = \myint \textrm{d}\textbf{r} \textrm{d}\textbf{r}' \textrm{d}\textbf{r}'' \frac{\psi_m(\textbf{r})\psi_{n}^*(\textbf{r}') \epsilon^{-1}(\textbf{r},\textbf{r}'';\omega=0){\psi_{m}^*}_{''}(\textbf{r})\psi_{n''}(\textbf{r}')}{\|\textbf{r}''-\textbf{r}'\|}\label{Screening}
%%\end{equation}
%%\begin{equation}
\W_{SS''} = \myint \textrm{d}\textbf{r} \textrm{d}\textbf{r}'  \frac{\psi_m(\textbf{r})\psi_{n}^*(\textbf{r}') \epsilon^{-1}(\textbf{r},\textbf{r}';\omega=0){\psi_{m}^*}_{''}(\textbf{r})\psi_{n''}(\textbf{r}')}{\|\textbf{r}-\textbf{r}'\|}\label{Screening}
\end{equation}
where $\epsilon^{-1}(\textbf{r},\textbf{r}';\omega)$ is approximated by a dielectric constant of four, that is, \(\epsilon^{-1}(\textbf{r},\textbf{r}';\omega) \approx \epsilon_\infty^{-1} \approx$ \sfrac{1}{4}, within the range-separation of the hybrid xc-functional.  
%where $\epsilon^{-1}(\textbf{r},\textbf{r}';\omega)$ is the dielectric function calculated within the random phase approximation (RPA).  
%%where $\epsilon^{-1}(\textbf{r},\textbf{r}';\omega)$ in reciprocal space is
%%\begin{equation}
%%\epsilon^{-1}_{\textbf{G}\textbf{G}'} = \sfrac{1}{4}\left[1 - e^{-\frac{\pi^2}{\mu^2}\|\textbf{G} + \Delta\textbf{k}\|^2}\right].
%%\end{equation}

Substituting (\ref{chi0}) into (\ref{BSE}), we can reformulate (\ref{BSE}) into an effective Hamiltonian eigenvalue problem
\begin{equation}
\Hmat_{SS'} \equiv (\varepsilon_n - \varepsilon_m)\delta_{n n'}\delta_{mm'} - (f_n - f_m)\left[\V_{SS'} - \W_{SS'}\right]
\end{equation}
with eigenvalues $E_{\lambda}$ and eigenvectors $A^S_{\lambda}$ from
\begin{equation}
\Hmat_{SS'} A_{\lambda}^S = E_{\lambda} A_{\lambda}^S.
\end{equation}
Here, $E_{\lambda}$ and $A^S_{\lambda}$ are the exciton energies and coefficients in electron-hole space.  The $\lambda$th exciton can then be expressed in terms of $A_{\lambda}^S$ and $\psi_n$ as
\begin{equation}
\Psi_{\lambda}(\textbf{r}_e,\textbf{r}_h) = \sum_{nm} A_{\lambda}^{nm} \psi_{n}(\textbf{r}_e)\psi_m(\textbf{r}_h)\label{Psi}
\end{equation}
where $\textbf{r}_e$ and $\textbf{r}_h$ are coordinates associated with the electron and hole, respectively.

To quantify the spatial distribution of the electron and hole of the $\lambda$th exciton, it is convenient to invoke the electron and hole densities, $\rho_e(\textbf{r}_e)$ and $\rho_h(\textbf{r}_h)$, which are obtained by integrating \eqref{Psi} with respect to the hole and electron coordinates, respectively\cite{Livia2014PSSB}
\begin{eqnarray}
\rho_e(\textbf{r}_e) &=& \myint \textrm{d}\textbf{r}_h\Psi_{\lambda}(\textbf{r}_e,\textbf{r}_h)\Psi_{\lambda}^*(\textbf{r}_e,\textbf{r}_h)\nonumber\\
&=& \sum_{nm}\sum_{n'}A_{\lambda}^{nm}{A_{\lambda}^{n' m}}^*\psi_n(\textbf{r}_e)\psi_{n'}^*(\textbf{r}_e).
\end{eqnarray}
Since
\begin{equation}
\myint \textrm{d}\textbf{r}_e A_{\lambda}^{nm}{A_{\lambda}^{n' m}}^*\psi_n(\textbf{r}_e)\psi_{n'}^*(\textbf{r}_e) = \left|A_{\lambda}^{nm}\right|^2\delta_{n n'}
\end{equation}
and by construction
\begin{equation}
\sum_{nm}\left|A_{\lambda}^{nm}\right|^2 \equiv 1,
\end{equation}
we ensure
\begin{equation}
\myint\textrm{d}\textbf{r}_e\rho_e(\textbf{r}_e) = 1.
\end{equation}
However, $A_{\lambda}^{nm}{A_{\lambda}^{n' m}}^*\psi_n(\textbf{r}_e)\psi_{n'}^*(\textbf{r}_e)$ is not necessarily zero for $n \neq n'$.  

The exciton's electron/hole densities, $\rho_{e/h}(\textbf{r}_{e/h})$, describe the spatial distribution of the electron/hole ``averaged'' over the hole/electron distribution.  To differentiate between Frenkel\cite{FrenkelExcitons} and Wannier\cite{WannierExcitons} excitons, one would need to calculate the two-point excitonic wave function $\Psi_\lambda(\textbf{r}_e,\textbf{r}_h)$, fix the position of the electron or hole, and unfold the $k$-point dependent wave function into real space. More precisely, including explicitly the $k$-point dependence, Eq \eqref{Psi} becomes
\begin{equation}
\Psi_{\lambda}(\textbf{r}_e,\textbf{r}_h) = \sum_{\textbf{k}}\sum_{nm} A_{\lambda}^{nm,\textbf{k}} \psi_{n,\textbf{k}}(\textbf{r}_e)e^{i\textbf{k}\cdot\textbf{r}_e}\psi_{m,\textbf{k}}(\textbf{r}_h)e^{i\textbf{k}\cdot\textbf{r}_h}\label{Psik}.
\end{equation}
This would require a significantly larger $k$-point sampling to probe whether the excitonic wave functions decay over several nanometers.  Further, the unit cells we have employed have too few layers to probe excitons with electron density in the bulk, and differentiate between bulk, surface, and bulk resonant levels.  For these reasons, we focus herein on the exciton's electron and hole densities, $\rho_{e}(\textbf{r}_e)$ and $\rho_{h}(\textbf{r}_h)$, respectively.

\section{RESULTS AND DISCUSSION}

\subsection{Overlayer Adsorption}\label{Subsect:Adsorption}
We begin by considering the most stable catechol overlayers on TiO$_2$(110) from the literature\cite{Catechol}.  Catechol consists of a benzene ring with two adjacent OH anchoring groups, which adsorbs on \Ticus{} sites of the rutile TiO$_2$(110) surface in a bidentate configuration\cite{Diebold,Catechol}.   From previous work\cite{Diebold,Catechol}, we know from comparing STM, UPS, and IPES experiments to $G_0W_0$ and HSE06 DFT calculations that the interface has a 1 ML $1\times4$ coverage.  We also consider the \sfrac{2}{3} ML $1\times3$ coverage to probe the dependence of the spectra on coverage and tilting of the molecule.

As shown in Figure~\ref{Structures:fgr},  upon adsorption the OH anchoring groups may dissociate, that is, deprotonate, with a charge transfer of $\sim-0.4 e$, indicated by arrows, accompanying the proton transfer to the adjacent bridging O atom (O$_{\textrm{br}}$) of the TiO$_2$(110) surface.  At 1 ML coverage, besides interfacial hydrogen bonds from the surface's hydrogenated bridging oxygens  (O$_{\textrm{br}}$H--C and O$_{\textrm{br}}$H--O), the overlayer, when not fully dissociated, is stabilized by intermolecular hydrogen bonds (OH--O) between neighboring catechol molecules, as depicted in Figure~\ref{Structures:fgr}.  For this reason, the 1 ML half (\sfrac{1}{2}D \sfrac{1}{2}D), mixed (D \sfrac{1}{2}D), and fully (D D) dissociated and \sfrac{2}{3} ML fully dissociated (D) catechol overlayers' adsorption energies are all within 0.1 eV, that is, the accuracy of DFT\cite{Catechol,CatecholCorrection}.  

In Table~\ref{Table1},
\begin{table}
%\vspace{-15.5mm}
\vspace{-19mm}
\noindent{\color{StyleColor}{\rule{\columnwidth}{1.0pt}}}
\vspace{14.5mm}
%\vspace{10.5mm}
\caption{\textrm{\bf Adsorption Energies \textit{E}$_{\textrm{ads}}$ in Electronvolts from PBE\cite{Catechol,CatecholCorrection}, HSE06, and vdW-DF DFT for 1 ML 1$\boldsymbol{\times}$4 and \sfrac{2}{3} ML 1$\boldsymbol{\times}$3 Half (\sfrac{1}{2}D) and Fully (D) Dissociated Catechol on TiO$_{\text{2}}$(110) with Tilting Angle $\boldsymbol{\theta}$ in Degrees and Number of Intermolecular (OH--O) and Interfacial (O$_{\textrm{br}}$H--C and O$_{\textrm{br}}$H--O) Bonds per Unit Cell }}\label{Table1}
%\tiny
\footnotesize
\begin{tabular}{@{}l@{}c@{}c@{}c@{}c@{}c@{}c@{}l@{}}
\multicolumn{8}{>{\columncolor[gray]{0.9}}c}{ }\\[-1mm]
\multicolumn{1}{>{\columncolor[gray]{.9}}c}{structure} & 
\multicolumn{1}{@{}>{\columncolor[gray]{.9}}c@{}}{$\theta$ (deg.)} & 
\multicolumn{3}{>{\columncolor[gray]{0.9}}c}{\underline{bonds/unit cell}}&
\multicolumn{3}{>{\columncolor[gray]{.9}}c}{\underline{$E_{\textrm{ads}}$ (eV)}}\\
\multicolumn{1}{>{\columncolor[gray]{0.9}}c}{ } &
\multicolumn{1}{>{\columncolor[gray]{0.9}}c}{} &
\multicolumn{1}{@{}>{\columncolor[gray]{.9}}c}{OH--O}& 
\multicolumn{1}{@{}>{\columncolor[gray]{.9}}c@{}}{O$_{\textrm{br}}$H--C} & 
\multicolumn{1}{>{\columncolor[gray]{.9}}c@{}}{O$_{\textrm{br}}$H--O} &
\multicolumn{1}{>{\columncolor[gray]{0.9}}c}{PBE}&
\multicolumn{1}{>{\columncolor[gray]{0.9}}c}{HSE06}&
\multicolumn{1}{@{}>{\columncolor[gray]{0.9}}l}{vdW-DF}
\\[0.5mm]

%% 1 ML Tilted Structures
%% Half Dissociated
% HD_HD_2OH2_pi
1 ML \sfrac{1}{2}D \sfrac{1}{2}D & %$\veebar$ &
 56, 49 & % 55.71, 49.40
2 & %0 & 
2 & 0 &
%-1217.4235 eV % PBE
$-0.614^a$&%&-0.645\\ % PBE
%-1607.284 (2x4) % HSE06
%% i2basque/m34L_plain_27V_shift_nbands320_HSE/HSE
% -288.71268 (2x1) (6.497 x 2.958) % Clean Surface % HSE06
%-112.1644  % Catechol % HSE06
$-0.944$&%\\% -0.94392 % HSE06
$-2.176$\\ % vdW-DF (single-point)
%$----$\\ % vdW-DF (relaxed)

%% Mixed Half and Fully Dissociated
% FD_HD_4OH
1 ML D \sfrac{1}{2}D & %$\veebar$ &
 56, 48 & % 55.92, 48.00
1 & %0 & 
1 & 2 &
% -1217.7072 % PBE
$-0.685^a$&%\\% & -0.697\\ % PBE
% -1607.6142 % HSE06
$-1.026$&%\\%-1.0264700000000175 % JSE
$-2.273$\\ % vdW-DF (single-point)
%$---$\\ % vdW-DF (relaxed)

%% Fuly Dissociated
1 ML D D & %$\veebar$ &
 56, 49 & % 55.81, 48.53
0 & %0 & 
2 & 2 &
% -1217.5737 % PBE
$-0.652^a$&%\\% & -0.665 % PBE
% -1607.5432 % HSE06
$-1.009$&%\\%-1.00872 % HSE06
$-2.228$\\ % vdW-DF (single-point)
%$-2.202$\\ % vdW-DF (relaxed)
%% FD_tilt / DISS_catechol_1x3_4L_C2h_tiltedplus_done
\sfrac{2}{3} ML D & %$\angle$ &
 \multicolumn{1}{c}{67} & % 67.40
0 & %0 & 
0 & 2 &
% -823.34143 % PBE
%-1.495$^a$& !!! Wrong!
$-0.748^b$&%-0.7476739999999751
% -1092.703 (2x3) (6.497 x 8.874) % HSE06
%% i2basque/m34L_plain_27V_shift_nbands320_HSE/HSE
% -288.71268 (2x1) (6.497 x 2.958) % Clean Surface % HSE06
%-112.1644  % Catechol % HSE06
% (-1092.703 - -1090.46684)/2
$-1.118$&%-1.11808%-2.236/2 %-2.23616/2
$-2.128$%\\ vdW-DF (single-point)
%$----$\\ % vdW-DF (relaxed)
\\%\hline
\multicolumn{8}{p{0.95\columnwidth}}{\tiny$^a$Reference~\citenum{Catechol}\nocite{Catechol} . $^b$Reference~\citenum{CatecholCorrection}\nocite{CatecholCorrection}.}
\end{tabular}
\noindent{\color{StyleColor}{\rule{\columnwidth}{1.0pt}}}
\end{table}
 we compare PBE DFT\cite{Catechol,CatecholCorrection}, HSE06 DFT, and vdW-DF DFT adsorption energies for the most stable catechol overlayers on TiO$_2$(110).  HSE06 DFT yields more stable absorption energies than PBE DFT by a constant shift of $\sim0.35$~eV.  The inclusion of van der Waals interactions (vdW-DF DFT) strengthens the adsorption energies for 1 ML catechol overlayers by a constant shift of $\sim1.58$~eV relative to PBE DFT.  However, at the lower \sfrac{2}{3} ML coverage where catechol is more upright ($\theta = 67^\circ$), the van der Waals contribution to the adsorption energy is reduced by 0.2 eV, inverting the relative stabilities of the 1 ML and \sfrac{2}{3} ML coverages.  This result is consistent with the experimental observation of 1 ML catechol overlayers on rutile TiO$_2$(110)\cite{Diebold, Catechol}. However, in all cases the adsorption energy per molecule is nearly isoenergetic for the half, mixed, fully dissociated 1 ML and dissociated \sfrac{2}{3} ML coverages.   Furthermore, relaxation of the overlayer's structure at the vdW-DF DFT level leaves both the structure and adsorption energies unchanged up to the accuracy of DFT.  This justifies our use of the PBE DFT geometries from ref~\citenum{Catechol}.

\subsection{Interfacial Level Alignment}\label{Subsect:Alignment}

For catechol on TiO$_2$(110), the relevant levels for absorption of the solar spectrum are from catechol's mid-gap HOMO and HOMO$-1$\cite{Diebold,Catechol} to the unoccupied Ti 3$d$ levels with $t_{2g}$-like symmetry $\{d_{xy},d_{xz},d_{y z}\}$\cite{DuncanTiO2,Selloni2PP,PetekPRB2015} of the substrate as the LUMO and LUMO$+1$ levels of catechol are significantly higher in energy\cite{Catechol}.  In fact, intramolecular HOMO$\rightarrow$LUMO transitions of the adsorbed molecular layer are significantly above the band gap of rutile TiO$_2$\cite{Catechol}.

An accurate description of the optical absorption spectra in the visible of the catechol--TiO$_2$(110) interface thus requires an accurate description of the interfacial level alignment of catechol's mid-gap HOMO and HOMO$-1$ and the unoccupied Ti 3$d$ levels of the substrate.  To quantify the accuracy of the calculated level alignment, one may compare the KS energy separation between catechol's HOMO and the substrate's CBM, $\Delta\varepsilon_{\textrm{HOMO}\rightarrow\textrm{CBM}} = \varepsilon_{\textrm{CBM}} - \varepsilon_{\textrm{HOMO}}$ with the binding energy of catechol's HOMO obtained from UPS experiments, as provided in Table~\ref{Table2}.
\begin{table}
%%\vspace{-15.5mm}
%\vspace{-19.mm}
%\noindent{\color{StyleColor}{\rule{\columnwidth}{1.0pt}}}
%%\vspace{9.5mm}
%\vspace{14.5mm}
\caption{\textrm{\bf HOMO$\boldsymbol{\rightarrow}$CBM KS Energy Differences $\boldsymbol{\Delta\varepsilon}_{\textrm{HOMO}\boldsymbol{\rightarrow}\textrm{CBM}}$, From PBE $\textit{G}_{\textrm{0}}\textit{W}_{\textrm{0}}$ and HSE06 DFT and the Lowest Energy Exciton's Energy \textit{E}$_{\textrm{(i)}}$ and Binding Energy \textit{E}$_{\textit{b}} = \textit{E}_{\textrm{(i)}} \boldsymbol{-} \boldsymbol{\Delta\varepsilon}_{\textrm{HOMO}\boldsymbol{\rightarrow}\textrm{CBM}}$ in Electronvolts and Intensity \textit{I}$_{\textrm{(i)}}$ From HSE06 BSE for 1 ML 1$\boldsymbol{\times}$4 and \sfrac{2}{3} ML 1$\boldsymbol{\times}$3 Half (\sfrac{1}{2}D) and Fully (D) Dissociated Catechol on TiO$_{\text{2}}$(110)}}\label{Table2}
%\tiny
\footnotesize
\begin{tabular}{@{}lccccc@{}}
\multicolumn{6}{>{\columncolor[gray]{0.9}}c}{ }\\[-1mm]
\multicolumn{1}{>{\columncolor[gray]{.9}}c}{structure} & 
\multicolumn{2}{>{\columncolor[gray]{.9}}c}{\underline{$\Delta\varepsilon_{\textrm{HOMO}\rightarrow\textrm{CBM}}$ (eV)}} &
\multicolumn{1}{@{}>{\columncolor[gray]{.9}}l}{$E_{\textrm{(i)}}$ (eV)}&
\multicolumn{1}{>{\columncolor[gray]{.9}}l}{$E_{\textit{b}}$ (eV)}&
\multicolumn{1}{>{\columncolor[gray]{.9}}c}{$I_{\textrm{(i)}}$}\\
\multicolumn{1}{>{\columncolor[gray]{.9}}c}{} & 
\multicolumn{1}{>{\columncolor[gray]{.9}}l}{PBE $G_0W_0$} & 
\multicolumn{1}{>{\columncolor[gray]{.9}}r@{}}{HSE06 DFT} & 
\multicolumn{3}{>{\columncolor[gray]{.9}}c}{\underline{HSE06 BSE}}
\\[0.5mm]
%% 1 ML Tilted Structures
%% Half Dissociated
% HD_HD_2OH2_pi
1 ML \sfrac{1}{2}D \sfrac{1}{2}D & %$\veebar$ &
3.926$^a$&% 3.92637 PBE G0W0
2.544& %2.544255 &% -0.313959--2.858214 (band 469 - band 468) 
%2.53176737 &0.45248063 
2.532 &
0.012 &
 0.452
\\

%% Mixed Half and Fully Dissociated
% FD_HD_4OH
1 ML D \sfrac{1}{2}D & %$\veebar$ &
3.441$^a$& % 3.440582 PBE G0W0
2.061& %2.060719& %-0.690121--2.750840 (band 469 - band 468)
%2.04470825 & 20.12530138
2.045 &
0.016 &
 20.125\hspace{1.5mm}
\\

%% Fuly Dissociated
1 ML D D & %$\veebar$ &
3.307$^a$& % 3.307188 PBE G0W0
1.942& %1.941704 & %-0.735257--2.676961
%1.93900108 & 1.31496043
1.939 &
0.003 &
 1.315
\\
%% FD_tilt / DISS_catechol_1x3_4L_C2h_tiltedplus_done
\sfrac{2}{3} ML D & 
%1.8458&%-1.1147--2.9605 (band 331 - band 330)
3.269$^{\hspace{1mm}}$&%3.269142 PBE G0W0
1.846&%1.84576&% -1.114716 - -2.960476
%1.84200 & 7.29737691
1.842 &
0.004 &
 7.297
\\
experiment & \multicolumn{2}{c}{2.4$^b$}\\
\multicolumn{6}{p{0.95\columnwidth}}{\tiny$^a$Reference~\citenum{Catechol}\nocite{Catechol}.  $^b$UPS binding energy relative to the Fermi level from ref~\citenum{Diebold}.}
\end{tabular}
\noindent{\color{StyleColor}{\rule{\columnwidth}{1.0pt}}}
\end{table}
Since the measured Fermi  level of rutile TiO$_2$ is approximately 0.1~eV below the CBM\cite{TiO2Fermi,TiO2FermiAono,TiO2FermiYamakata},  the energy separation between catechol's HOMO and the CBM of the TiO$_2$(110) surface from UPS measurements is $\Delta\varepsilon_{\textrm{HOMO}\rightarrow\textrm{CBM}} \approx 2.4 + 0.1 = 2.5$~eV\cite{Diebold}.

As shown in Table~\ref{Table2}, $\Delta\varepsilon_{\textrm{HOMO}\rightarrow\textrm{CBM}}$ from HSE06 DFT agrees semi-quantitatively with the binding energy measured with UPS for catechol on TiO$_2$(110)\cite{Diebold}.  Conversely, PBE $G_0W_0$ overestimates $\Delta\varepsilon_{\textrm{HOMO}\rightarrow\textrm{CBM}}$ by $\sim 1$~eV. This is consistent with the overestimation of the electronic gap for the TiO$_2$(110) surface by $\sim1$~eV with PBE $G_0W_0$\cite{MiganiLong}.  This motivates our use of HSE06 DFT eigenvalues as input to our BSE calculations.

\begin{figure}%[!htb]
\noindent{\color{StyleColor}{\rule{\columnwidth}{1pt}}}
\includegraphics[width=\columnwidth]{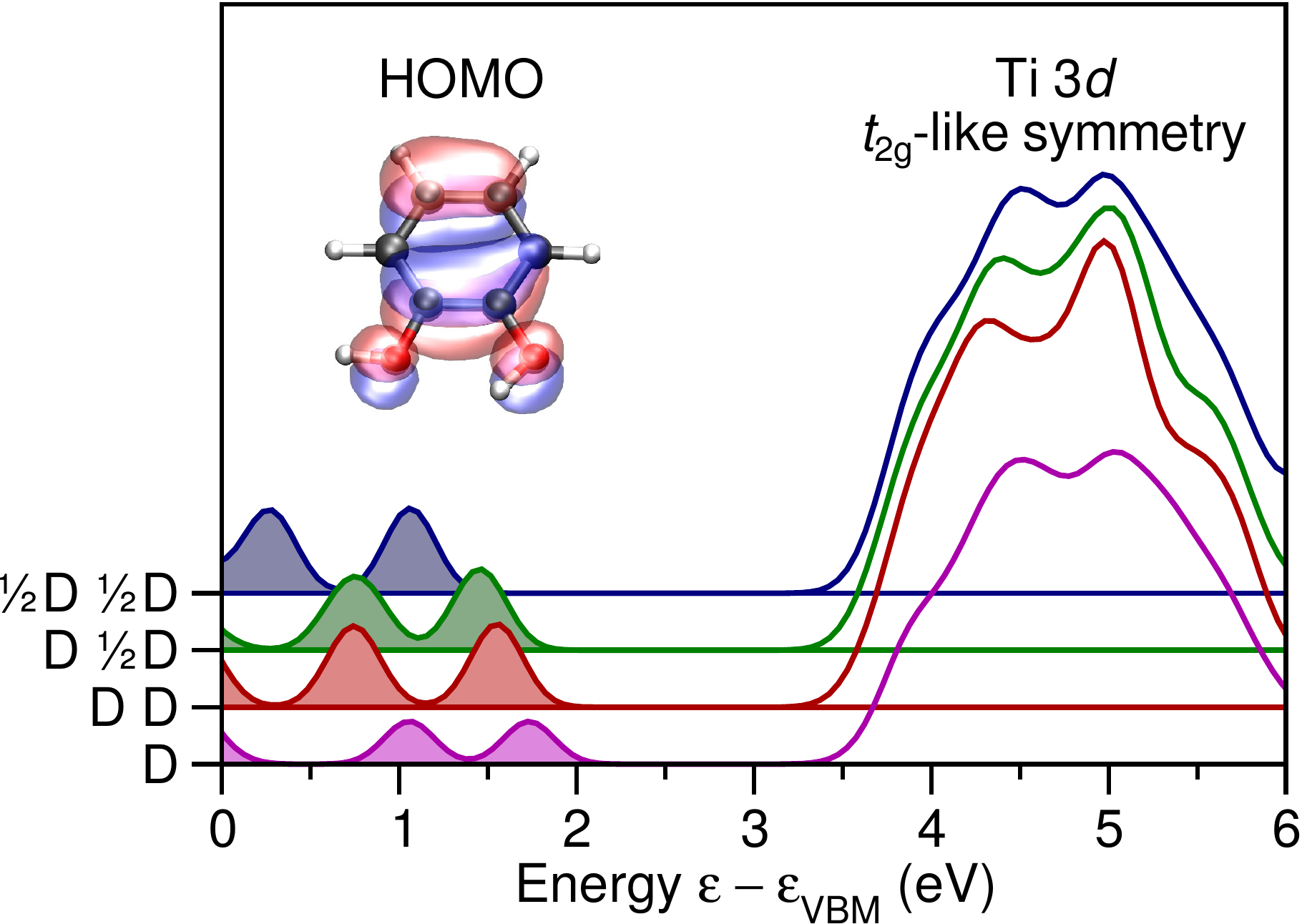}
\caption{HSE06 DFT total DOS for 1 ML half (\sfrac{1}{2}D \sfrac{1}{2}D, blue), mixed (D \sfrac{1}{2}D, green), and fully (D D, red) dissociated and \sfrac{2}{3} ML fully (D, violet) dissociated catechol on TiO$_2$(110).  Energies are relative to the VBM  ($\varepsilon_{\textrm{VBM}}$) of clean TiO$_2$(110).  Filling denotes occupation. The HOMO of isolated catechol is shown as an inset.}\label{fgr:DOS}
\noindent{\color{StyleColor}{\rule{\columnwidth}{1pt}}}
\end{figure}

In Figure~\ref{fgr:DOS},
 we compare the HSE06 DOS for the 1 ML half (\sfrac{1}{2}D \sfrac{1}{2}D), mixed (D \sfrac{1}{2}D),  and fully (D D) dissociated and \sfrac{2}{3} ML dissociated (D) catechol on TiO$_2$(110).  In each case, there are two distinct peaks outside the clean surface's VB region.  These two peaks are associated with the HOMO$-1$ and HOMO levels of catechol\cite{Catechol}.  As a reference, the HOMO of gas phase catechol, which has $\pi$ character, is depicted in Figure~\ref{fgr:DOS}. 

\begin{figure*}[!t]
%\noindent{\color{StyleColor}{\rule{\textwidth}{1pt}}}
\includegraphics[width=\textwidth]{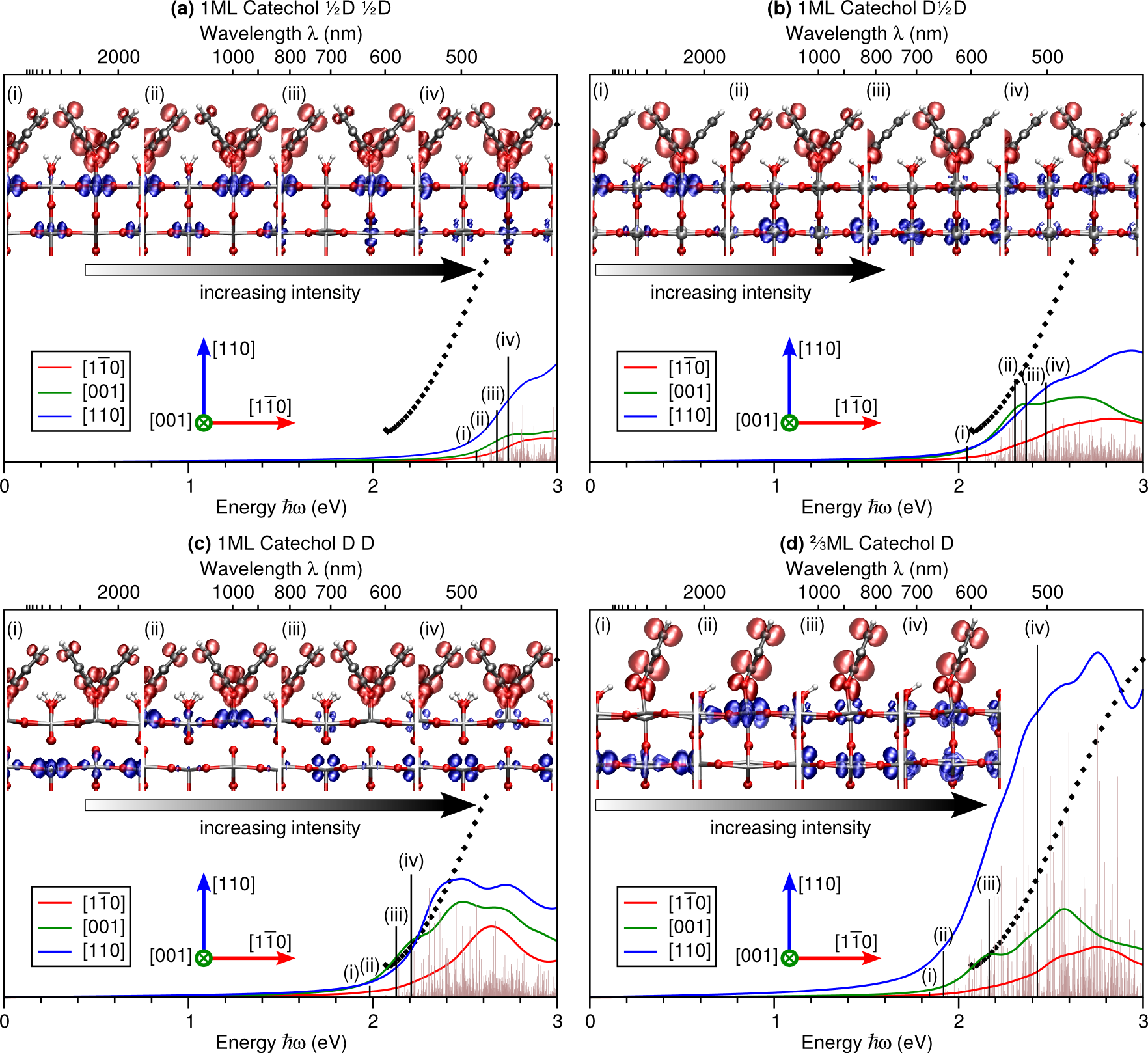}
\caption{HSE06 BSE optical absorption spectra  for light polarized along [1$\overline{1}$0] (red), [001] (green), and [110] (blue) directions of 1 ML \textbf{(a)} half (\sfrac{1}{2}D \sfrac{1}{2}D), \textbf{(b)} mixed (D \sfrac{1}{2}D), and \textbf{(c)} fully (D D) dissociated  and \textbf{(d)} \sfrac{2}{3} ML fully dissociated catechol on TiO$_2$(110).   Vertical lines denote the energies and intensities of individual excitons (brown).  The electron (blue) and hole (red) densities, $\rho_e$ and $\rho_h$, of four representative excitons (i--iv) chosen in order of increasing intensity with energy (black vertical lines) are shown as insets. Measured optical absorption of catechol on anatase TiO$_2$ nanoparticles (black diamonds) is taken from ref.~\citenum{CatecholTiO2NPJPCC2012}.}\label{fgr:StructBSE}
\noindent{\color{StyleColor}{\rule{\textwidth}{1pt}}}
\end{figure*}

 Figure~\ref{fgr:DOS} shows that the HOMO and HOMO$-1$ are pinned to each other, and shift up in energy with deprotonation of adsorbed catechol.  This deprotonation of the OH anchoring group induces a charge transfer of $\sim-0.4e$ to the substrate.  As charge is removed from the molecule, the HOMO and HOMO$-1$ are destabilized.  This is because the molecule's ability to screen the HOMO and HOMO$-1$ levels is reduced as charge is transferred to the substrate\cite{OurJACS}.  Consequently, the energy separation between the HOMO and CBM, $\Delta\varepsilon_{\textrm{HOMO}\rightarrow\textrm{CBM}} = \varepsilon_{\textrm{CBM}} - \varepsilon_{\textrm{HOMO}}$, decreases with dissociation by $\sim 0.6$~eV, as provided in Table~\ref{Table2}.  Moreover, $\Delta\varepsilon_{\textrm{HOMO}\rightarrow\textrm{CBM}}$ decreases by a further 0.1~eV as the coverage decreases from 1 ML to \sfrac{2}{3} ML.  This already suggests fully dissociated catechol overlayers at lower coverage may have greater photovoltaic efficiencies.

\subsection{Sub-Band Gap Optical Absorption}\label{Subsect:Absorption}

Optical absorption above the substrate band gap of 3.3~eV\cite{TiO2BandGap} will be dominated by supra-band gap transitions within the substrate.  Here, we are focused on the photovoltaic efficiency of the dye-substrate interface.  
This means we may restrict consideration to sub-band gap transitions which are below the optical gap of rutile TiO$_2$ at 3.0~eV\cite{RutileOpticalGap,PhysRevB.18.5606}.  Moreover, these transitions are most relevant for absorption of solar irradiation.

For systems with weak electron-hole binding, the onset of the optical absorption spectra may already be well described within an independent-particle (IP) picture\cite{KresseRPA,LiviaMilan}.  Further, excitons composed primarily of a single transition between KS levels would also be well described at the IP level.  As shown in Figure~S1 of the supporting information (SI), this is indeed the case for \sfrac{2}{3}~ML D catechol on TiO$_2$(110), although significant differences in relative intensities with the BSE spectra are observed for catechol monolayers on TiO$_2$(110).  However, the degree of electron-hole binding and mixing of the transitions that compose the relevant excitons is not known \textit{a priori}.  As a result, the reliability of the IP spectra can only be assessed after obtaining the BSE spectra.

In Figure~\ref{fgr:StructBSE} we show the angular resolved optical absorption spectra along the $[1\overline{1}0]$, $[001]$, and $[110]$ directions for 1 ML \sfrac{1}{2}D \sfrac{1}{2}D, D \sfrac{1}{2}D, D D, and \sfrac{2}{3} ML D catechol on TiO$_2$(110).  We also show the spatial distribution of four representative excitons (i--iv) that have been chosen in order of increasing intensity with energy, as depicted schematically using arrows in Figure~\ref{fgr:StructBSE}. The computed spectra include direct transitions from catechol's HOMO and HOMO$-1$ to the unoccupied Ti 3$d$ conduction levels with $t_{2g}$-like symmetry $\{d_{xy},d_{xz},d_{y z}\}$.  The last $N_{\textrm{cat}}$ occupied levels are bonding and antibonding levels composed of the HOMOs of the catechol molecules within our unit cells\cite{Catechol}.  In Table~\ref{Table3} 
\begin{table}[!htb]
\vspace{-22mm}
\noindent{\color{StyleColor}{\rule{\columnwidth}{1.0pt}}}
\vspace{17.5mm}
\caption{\textrm{\bf Energy \textit{E}$_{\boldsymbol{\lambda}}$ in Electronvolts, Intensity \textit{I}$_{\boldsymbol{\lambda}}$, and Transitions From Catechol's HOMO to TiO$_{\text{2}}$(110)'s CB $\textit{m}\boldsymbol{\rightarrow}\textit{n}$ Relative to the Number of Occupied Levels $\textit{N}_\textrm{occ}$ at Given \textit{k}-points with Amplitudes $\boldsymbol{|}\textit{A}_{\boldsymbol{\lambda}}^{\textit{nm}}\boldsymbol{|}^\textrm{2}$ That Form  the Four Excitons (i--iv) Shown in Figure~\ref{fgr:StructBSE} for 1 ML 1$\boldsymbol{\times}$4 (\textit{m} $\boldsymbol{-}$ \textit{N}$_{\textbf{occ}}$ = 0, $\boldsymbol{-}$1, $\boldsymbol{-}$2, $\boldsymbol{-}$3) and \sfrac{2}{3} ML 1$\boldsymbol{\times}$3 (\textit{m} $\boldsymbol{-}$ \textit{N}$_{\textbf{occ}}$ = 0, $\boldsymbol{-}$1) Half (\sfrac{1}{2}D) and Fully (D) Dissociated Catechol on TiO$_{\text{2}}$(110)}}\label{Table3}
%\tiny
\footnotesize
\begin{tabular}{@{}c@{}c@{}c@{}r@{ $\rightarrow$ }l@{}c@{}c}
\multicolumn{7}{>{\columncolor[gray]{0.9}}c}{ }\\[-1mm]
\multicolumn{3}{>{\columncolor[gray]{.9}}c}{\underline{exciton}} & 
\multicolumn{4}{>{\columncolor[gray]{.9}}c}{\underline{transition}} %&
%\multicolumn{2}{>{\columncolor[gray]{.9}}c}{}
\\
\multicolumn{1}{>{\columncolor[gray]{.9}}c}{$\lambda$} &
\multicolumn{1}{@{}>{\columncolor[gray]{.9}}c@{}}{$E_\lambda$ (eV)} &
\multicolumn{1}{>{\columncolor[gray]{.9}}c}{$I_\lambda$ (arb. units)} &
\multicolumn{1}{>{\columncolor[gray]{.9}}c}{$m - N_{\textrm{occ}}$} &
\multicolumn{1}{@{}>{\columncolor[gray]{.9}}c}{$n - N_{\textrm{occ}}-1$} &
\multicolumn{1}{>{\columncolor[gray]{.9}}c}{$k$-point}&
\multicolumn{1}{>{\columncolor[gray]{.9}}c}{$|A_\lambda^{nm}|^2$}\\%[0.5mm]
% (a) half dissociated (1/2D 1/2D) 1ML catechol overlayer
\multicolumn{7}{>{\columncolor[gray]{.9}}c}{1 ML catechol \sfrac{1}{2}D \sfrac{1}{2}D}\\[0.5mm]
%1 ML \sfrac{1}{2}D \sfrac{1}{2}D &
 (i) & 2.532& 0.452 & $-1$&$0$, $+2$ & $\Gamma$ & 68\%\\
% exciton (i) $-1\rightarrow$/$+2$ at $\Gamma$ (68\%)
%1 ML \sfrac{1}{2}D \sfrac{1}{2}D& 
&&    & $0$&$+1$, $+3$ & $\Gamma$ & 20\%\\
% exciton (i) $\rightarrow$$+1$/$+3$ at $\Gamma$ (20\%)
%1 ML \sfrac{1}{2}D \sfrac{1}{2}D &
 (ii)& 2.562& 14.192\hspace{1mm} & $-2$&$0$, $+2$ & $\Gamma$ & 68\%\\
% exciton (ii) $-2\rightarrow0$/$+2$ at $\Gamma$ (68\%)
%1 ML \sfrac{1}{2}D \sfrac{1}{2}D &
&&  & $-3$&$+1$, $+3$ & $\Gamma$ & 18\%\\
% exciton (ii) $-3\rightarrow$$+1$/$+3$ at $\Gamma$ (18\%)
% exciton (ii) $-2\rightarrow$$+2$ at (0,$\pm$\sfrac{1}{3}, 0) (3\%)
% exciton (ii) $-3\rightarrow$$+3$ at (0,$\pm$\sfrac{1}{3}, 0) (3\%)
%1 ML \sfrac{1}{2}D \sfrac{1}{2}D &
 (iii)&2.672& 67.166\hspace{1mm}& $-3$&$+3$ & $\Gamma$, $\pm$\sfrac{2}{3}Y &15\%\\
% exciton (iii) $-3\rightarrow$$+3$ at $\Gamma$, (0, $\pm$\sfrac{1}{3}, 0) (15\%)
%1 ML \sfrac{1}{2}D \sfrac{1}{2}D &
&&  & $-2$&$+2$ & $\Gamma$, $\pm$\sfrac{2}{3}Y &17\%\\
% exciton (iii) $-2\rightarrow$$+2$ at $\Gamma$, (0, $\pm$\sfrac{1}{3}, 0) (17\%)
%1 ML \sfrac{1}{2}D \sfrac{1}{2}D &
&&  & $-3$&$0$ & $\pm$\sfrac{1}{2}X & 10\%\\
% exciton (iii) $-3\rightarrow$$0$ at ($\pm$\sfrac{1}{4}, 0, 0) (10\%)
%1 ML \sfrac{1}{2}D \sfrac{1}{2}D &
&&  & $0$&$+1$ & $\pm$\sfrac{1}{2}X& 6\%\\
% exciton (iii) $\rightarrow$$+1$ at ($\pm$\sfrac{1}{4}, 0, 0) (6\%)
%1 ML \sfrac{1}{2}D \sfrac{1}{2}D &
 (iv)&2.734& 136.457\hspace{2.5mm} & $-1$&$+2$ & $\pm$\sfrac{1}{2}X & 19\%\\
% exciton (iv) $-1\rightarrow$$+2$ at ($\pm$\sfrac{1}{4}, 0, 0) (19\%)
%1 ML \sfrac{1}{2}D \sfrac{1}{2}D &
&&  & $-3$, $-2$&$+3$ & $\pm$\sfrac{2}{3}Y & 17\%\\[0.5mm]
% exciton (iv) $-3$/$-2\rightarrow$$+3$ at (0,$\pm$\sfrac{1}{3},0) (17\%)
% (b) mixed dissociated (D 1/2D) 1 ML catechol overlayer 
\multicolumn{7}{>{\columncolor[gray]{0.9}}c}{ }\\[-1mm]
\multicolumn{7}{>{\columncolor[gray]{.9}}c}{1 ML catechol D \sfrac{1}{2}D}\\[0.5mm]
%1 ML D \sfrac{1}{2}D &
 (i)&2.045& 20.125\hspace{1mm} & $0$&$0$, $+3$& $\Gamma$ & 53\%\\
% exciton (i) $\rightarrow$/$+3$ at $\Gamma$ (53\%)
%1 ML D \sfrac{1}{2}D &
&& & $-1$&$+1$, $+2$& $\Gamma$ & 30\%\\
% exciton (i) $-1\rightarrow$$+1$/$+2$ at $\Gamma$ (30\%)
%1 ML D \sfrac{1}{2}D &
(ii)&2.305 & 107.222\hspace{2.5mm} & $0$&$+1$ & $\pm$\sfrac{1}{2}X$\pm$\sfrac{2}{3}Y &51\%\\
% exciton (ii) $\rightarrow$$+1$ at ($\pm$\sfrac{1}{4},$\pm$\sfrac{1}{3},0) (51\%)
%1 ML D \sfrac{1}{2}D &
&& & $-3$&$+1$ & $\pm$\sfrac{2}{3}Y& 12\%\\
% exciton (ii) $-3\rightarrow$$+1$ at (0, $\pm$\sfrac{1}{3}, 0) (12\%)
%1 ML D \sfrac{1}{2}D &
&& & $0$&$+2$ & $\pm$\sfrac{1}{2}X & 7\%\\
% exciton (ii) $\rightarrow$$+2$ at ($\pm$\sfrac{1}{4}, 0, 0) (7\%)
%1 ML D \sfrac{1}{2}D &
 (iii)&2.366 & 102.162\hspace{2.5mm} & $0$&$+5$ &$\pm$\sfrac{2}{3}Y &23\%\\
% exciton (iii) 0$\rightarrow$$+5$ at (0, $\pm$\sfrac{1}{3}, 0) (23\%)
%1 ML D \sfrac{1}{2}D &
&& & $-1$&$+5$ & $\Gamma$ &11\%\\
% exciton (iii) $-1\rightarrow$$+5$ at $\Gamma$ (11\%)
%1 ML D \sfrac{1}{2}D &
&& & $0$&$+4$ &$\pm$X &19\%\\
%1 ML D \sfrac{1}{2}D &
&& & $-1$&$+5$ &$\pm$X &10\%\\
% exciton (iii) $-1$/$\rightarrow$$+4$ at ($\pm$\sfrac{1}{2}, 0, 0) (29\%)
%1 ML D \sfrac{1}{2}D &
 (iv)&2.473& 102.949\hspace{2.5mm} & $-1$&$+7$ &$\pm$\sfrac{1}{2}X$\pm$\sfrac{2}{3}Y &34\%\\
% exciton (iv) $-1\rightarrow$$+7$ at ($\pm$\sfrac{1}{4},$\pm$\sfrac{1}{3}, 0) (34\%)
%1 ML D \sfrac{1}{2}D &
&& & $0$&$+10$ & $\pm$X$\pm$\sfrac{2}{3}Y &17\%\\
% exciton (iv) $\rightarrow$$+10$ at ($\pm$\sfrac{1}{2},$\pm$\sfrac{1}{3}, 0) (17\%)
%1 ML D \sfrac{1}{2}D &
&&  & $0$&$+7$ &$\mp$\sfrac{1}{2}X$\pm$\sfrac{2}{3}Y &5\%\\
% exciton (iv) $\rightarrow$$+7$ at ($\mp$\sfrac{1}{4},$\pm$\sfrac{1}{3}, 0) (5\%)
%1 ML D \sfrac{1}{2}D &
&&  & $0$&$+8$ & $\pm$\sfrac{1}{2}X$\pm$\sfrac{2}{3}Y &4\%\\[0.5mm]
% exciton (iv) $\rightarrow$$+8$ at ($\pm$\sfrac{1}{4},$\pm$\sfrac{1}{3}, 0) (4\%)
% (c) fully dissociated (D D) 1 ML catechol overlayer
\multicolumn{7}{>{\columncolor[gray]{0.9}}c}{ }\\[-1mm]
\multicolumn{7}{>{\columncolor[gray]{.9}}c}{1 ML catechol D D}\\[0.5mm]
%1 ML D D &
 (i)&1.939& 1.315 & $-1$&$0$ & $\Gamma$ &95\%\\
% exciton (i) $-1\rightarrow$$0$ at $\Gamma$ (95\%)
%1 ML D D &
&&  & $0$&$0$ & $\Gamma$ &4\%\\
% exciton (i) $\rightarrow$$0$ at $\Gamma$ (4\%)
%1 ML D D &
 (ii)&1.983& 14.925\hspace{1mm} & $0$&$+1$, $+3$ & $\Gamma$ & 52\%\\
% exciton (ii) $\rightarrow$$+1$/$+3$ at $\Gamma$ (52\%)
%1 ML D D &
&&  & $-1$&$+2$ & $\Gamma$ & 31\%\\
% exciton (ii) $-1\rightarrow$$+2$ at $\Gamma$ (31\%)
%1 ML D D &
 (iii)&2.127& 91.871\hspace{1mm}& $-1$&$+4$ & $\Gamma$ & 53\%\\
% exciton (iii) $-1\rightarrow$$+4$ at $\Gamma$ (53\%)
%1 ML D D &
&&  & $0$&$+2$ & $\pm$X$\pm$\sfrac{2}{3}Y &8\%\\ 
% exciton (iii) $\rightarrow$$+2$ ($\pm$\sfrac{1}{2},$\pm$\sfrac{1}{3},0) (8\%) 
%1 ML D D &
 (iv)&2.208& 158.836\hspace{2.5mm}& $-1$, $0$&$+4$ &$\pm$\sfrac{2}{3}Y &74\%\\
% exciton (iv) /$-1\rightarrow$$+4$ (0, $\pm$\sfrac{1}{3}, 0) (74\%)
%1 ML D D &
&& & $-1$, $0$&$+1$ &$\pm$\sfrac{1}{2}X$\pm$\sfrac{2}{3}Y & 4\%\\[0.5mm]
% exciton (iv) $-1$/$\rightarrow$$+1$ ($\pm$\sfrac{1}{4},$\pm$\sfrac{1}{3}, 0) (4\%)
% (d) dissociated \sfrac{2}{3} ML catechol overlayer
\multicolumn{7}{>{\columncolor[gray]{0.9}}c}{ }\\[-1mm]
\multicolumn{7}{>{\columncolor[gray]{.9}}c}{\sfrac{2}{3} ML catechol D}\\[0.5mm]
%\sfrac{2}{3} ML D &
 (i)&1.842& 7.297 & $0$&$0$& $\Gamma$ & 98\%\\
% exciton (i) $\rightarrow$$0$ at $\Gamma$ (98\%)
%\sfrac{2}{3} ML D &
 (ii)&1.917& 59.896\hspace{1.mm}& $-1$&$+2$ & $\Gamma$ & 42\%\\
%\sfrac{2}{3} ML D &
&&  & $-1$&$+2$ & $\pm$\sfrac{2}{3}Y & 2\%\\
% excition (ii) $-1\rightarrow$$+2$ at $\Gamma$, (42\%) (0,$\pm$\sfrac{1}{3},0) (2\%)
%\sfrac{2}{3} ML D &
&&  & $0$&$+3$ & $\Gamma$ & 37\%\\
%\sfrac{2}{3} ML D &
&&  & $0$&$+3$ & $\pm$\sfrac{2}{3}Y & 2\%\\
% excition (ii) $\rightarrow$$+3$ at $\Gamma$, (37\%) (0,$\pm$\sfrac{1}{3},0) (3\%)
%\sfrac{2}{3} ML D &
 (iii)&2.165& 127.158\hspace{2.5mm} & $-1$&$+4$ & $\Gamma$ &54\%\\
% exciton (iii) $-1\rightarrow$$+4$ at $\Gamma$, (54\%)
%\sfrac{2}{3} ML D &
&&  & $0$&$+6$ & $\Gamma$ & 10\%\\
% exciton (iii) $\rightarrow$$+6$ at $\Gamma$, (10\%)
%\sfrac{2}{3} ML D &
&&  & $0$&$+4$ & $\pm$\sfrac{2}{3}Y & 13\%\\
% exciton (iii) $\rightarrow$$+4$ (0, $\pm$\sfrac{1}{3}, 0) (13\%)
%\sfrac{2}{3} ML D &
 (iv)&2.427& 455.011\hspace{2.5mm} &$0$& $+8$ & $\pm$\sfrac{2}{3}Y & 90\%\\
% exciton (iv) $\rightarrow$$+8$ (0, $\pm$\sfrac{1}{3}, 0) (90\%)
\end{tabular}
\noindent{\color{StyleColor}{\rule{\columnwidth}{1.0pt}}}
\end{table}
these are referred to as $m-N_{\textrm{occ}} = 0, -1, -2, -3$ and $m-N_{\textrm{occ}} = 0, -1$ for 1 ML 1$\times$4 and \sfrac{2}{3} ML 1$\times$3 catechol overlayers, respectively.

The individual KS transitions that form the four excitons (i--iv) shown in Figure~\ref{fgr:StructBSE} are provided in Table~\ref{Table3}.  As these are charge transfer excitations, one might expect BSE calculations to underestimate their intensity \cite{CTexcitonsCatecholAnatase,LluisCT,ZwijenburgJCTC2014a,ZwijenburgJCTC2014b,CTHSE,NunziJCTC2015}.  However, as shown in Figure~\ref{fgr:StructBSE} and Table~\ref{Table2}, we find rather intense charge transfer excitons below 3 eV.  Their hole density is localized on catechol's HOMO, while the electron density is composed of Ti 3$d_{xy}$ orbitals. These transitions result in significant absorption in the visible region (400 -- 700 nm) for the fully dissociated overlayers at \sfrac{2}{3} and 1 ML coverages.   For half and mixed dissociated 1 ML overlayers, the HOMO and HOMO$-1$ are shifted lower in energy, resulting in a shift to higher energies of the absorption spectra. For  1 ML coverages, we find absorption along the [110] direction has similar intensities, while absorption in the surface plane increases in intensity with dissociation of the catechol overlayer (\textit{cf}.\ Figure~\ref{fgr:StructBSE}(a--c)).    

Overall, our spectra for the catechol overlayers are consistent with that measured for colloidal catechol on anatase TiO$_2$ nanoparticles\cite{GratzelExpCatecholTiO2NP,CatecholTiO2NPTJPCB2003,CatecholTiO2NPJPCC2012,CatecholTiO2NPReviewEES2014}.  As shown in Figure~\ref{fgr:StructBSE}, the measured spectrum has an onset at 600 nm, and a noticeable shoulder at 430 nm, in semi-quantitative agreement with our calculated optical absorption for catechol on rutile TiO$_2$(110).  Moreover, taking into account the $\sim 0.3$~eV lower direct electronic band gap of rutile compared to anatase TiO$_2$\cite{SunH2OAnatase}, the measured and calculated spectra would shift into near quantitative agreement for the most stable \sfrac{2}{3} ML catechol overlayer.  This is consistent with the red shift of the absorption spectrum calculated with PBE time dependent (TD)DFT in ref.~\citenum{SelloniCatecholTDDFT} for catechol on rutile compared to anatase TiO$_2$.

The spatial distribution of the lowest energy exciton, exciton (i), is shown for each catechol overlayer in Figure~\ref{fgr:StructBSE}.  For half \sfrac{1}{2}D \sfrac{1}{2}D and mixed D \sfrac{1}{2}D dissociated catechol monolayers, exciton (i) has electron density located predominantly on surface \Ticus{} sites, with some weight on subsurface Ti atoms, as shown in Figure~\ref{fgr:StructBSE}(a) and (b), respectively.  For fully dissociated 1 ML and \sfrac{2}{3} ML catechol overlayers, exciton (i) has electron density located almost entirely on subsurface Ti atoms, as shown in Figure~\ref{fgr:StructBSE}(c) and (d).  In each case, exciton (i) is composed of HOMO$\rightarrow$CBM Ti 3$d_{xy}$ transitions at $\Gamma$, as shown in Table~\ref{Table3}.  This suggests that fully deprotonating catechol's OH anchoring groups induces a larger spatial separation between the electron and hole normal to the surface.  The binding of exciton (i) is quite weak, $\Delta\varepsilon_{\textrm{HOMO}\rightarrow\textrm{CBM}} - E_{\textrm{(i)}} \ll 25$~meV, as seen from Table~\ref{Table2}. This means, at the BSE level, the onset of the spectra decreases with dissociation and reduction in coverage, due to the associated upshift of catechol's HOMO levels, shown in Figure~\ref{fgr:DOS}.  

Although the binding energies are very small (\textit{cf}.\ Table~\ref{Table2}), we find when the electron density is located on the surface Ti atoms (\textit{cf}.\ Figure~\ref{fgr:StructBSE}(a) and (b)), the binding energy is about four times greater than when the electron density  is located on subsurface Ti atoms (\textit{cf}.\ Figure~\ref{fgr:StructBSE}(c) and (d)).  This suggests, although electron and hole binding is greater when the electron density is located on the surface, separation of charge carriers is facile for catechol--TiO$_2$(110) interfaces.

The nature of the two different types of excitons with either smaller (\textit{cf.} Figure~\ref{fgr:StructBSE}(a) and (b)) or larger (\textit{cf.} Figure~\ref{fgr:StructBSE}(c) and (d)) separation normal to the surface between the charge carriers is in line with transient absorption spectra\cite{CatecholTiO2NPTJPCB2003}. These indicate that back-electron transfer dynamics, that is, recovery time dynamics, are bi-exponential, with shorter and longer components. The fast component  is in agreement with strong electronic coupling between charge carriers in these charge transfer states, and is attributed to the recovery time for electrons injected at Ti centers close to the adsorbate, e.g., \Ticus{} sites. The slow component is attributed to the recovery time for electrons at a larger distance from the adsorbate. These results point to the important effect of charge separation on recovery time dynamics\cite{CatecholTiO2NPTJPCB2003,CatecholTiO2NPJPCC2012}.

From Table~\ref{Table2}, we also find excitons (i) shown in Figure~\ref{fgr:StructBSE}(b) and (d) are an order of magnitude more intense than those shown in Figure~\ref{fgr:StructBSE}(a) and (c).  More importantly, by reducing the coverage for fully dissociated catechol from 1 ML to \sfrac{2}{3} ML, absorption normal to the surface increases significantly in intensity, while absorption in the surface plane is somewhat reduced.  This correlates with the catechol overlayer becoming more upright as the coverage is reduced (\textit{cf}.\ Table~\ref{Table1}), which should increase the dipole moment normal to the surface.  

For the half dissociated \sfrac{1}{2}D \sfrac{1}{2}D catechol monolayer, the first two excitons, (i) and (ii), of Figure~\ref{fgr:StructBSE}(a), are predominantly electronic transitions from HOMO$\rightarrow$CBM at $\Gamma$ (\textit{cf.} Table~\ref{Table3}).  Exciton (i) has hole density mostly on the HOMO of the more upright molecule, while exciton (ii) has the hole density mostly on the HOMO of the more tilted molecule\cite{Catechol}.  In both cases, the electron density is on the CBM, which has predominantly \Ticus{} 3$d_{xy}$ character, with some weight on the $d_{xy}$ orbitals of the Ti atoms below O$_{\textrm{br}}$ atoms\cite{SunH2OAnatase}.  Exciton (iii) has a hole density similar to exciton (ii), with the electron density again having predominantly  \Ticus{} 3$d_{xy}$ character, but with some weight on the $d_{xy}$ orbitals of the Ti atoms below \Ticus{} sites.   The latter is attributable to the inclusion of transitions at other $k$-points (\textit{cf}.\ Table~\ref{Table3}). The hole density of exciton (iv) is a rather even mixture between the HOMOs of the two molecules, while the electron density of exciton (iv) is a mixture of $d_{xy}$, $d_{xz}$, and $d_{y z}$ orbitals predominantly of \Ticus{} sites, with some weight on the Ti atoms of the sub-layer.  This exciton is composed of transitions at $k$-points away from $\Gamma$ (\textit{cf}.\ Table~\ref{Table3}). The resulting \Ticus{} 3$d$ electron density yields a greater overlap with the O 2$p_\pi$ levels of catechol's HOMO.  This may explain exciton (iv)'s greater intensity.

For the mixed D \sfrac{1}{2}D catechol monolayer, the hole density is predominantly on the more upright fully dissociated catechol molecules of the overlayer, as seen for excitons (i), (iii), and (iv) of Figure~\ref{fgr:StructBSE}(b).  Exciton (i) is predominantly a HOMO$\rightarrow$CBM transition at $\Gamma$, with electron density mostly \Ticus{} 3$d_{xy}$ in character.  The electron densities of excitons (ii) and (iii) are a mixture of $d_{xz}$ and $d_{y z}$ orbitals predominantly on subsurface Ti atoms.  These excitons are mostly composed of transitions away from $\Gamma$ (\textit{cf}.\ Table~\ref{Table3}). However, despite the significant spatial separation normal to the surface between the hole and electron densities which form excitons (ii) and (iii), their intensities are rather significant and similar.  Exciton (iv), which has a similar intensity, has electron density predominantly on surface Ti atoms, with a mixed $d_{xy}$, $d_{xz}$, and $d_{y z}$ character.  This exciton is composed of transitions from HOMO to higher energy CB levels mostly in the middle of the Brillouin zone (\textit{cf}.\ Table~\ref{Table3}).  

The fully dissociated catechol 1 ML and \sfrac{2}{3} ML overlayers have excitons (i--iv) with similar electron densities, as seen in Figure~\ref{fgr:StructBSE}(c) and (d).  For both coverages,  exciton (i) is composed of HOMO$\rightarrow$CBM transitions at $\Gamma$ (\textit{cf.} Table~\ref{Table3}), with electron densities predominantly $d_{xy}$ in character, on subsurface Ti atoms.  Similarly, for both coverages, exciton (ii) is composed of transitions mostly at $\Gamma$ from HOMO to higher energy CB levels (\textit{cf.} Table~\ref{Table3}), with electron densities predominantly $d_{xy}$ in character, but on surface Ti atoms. For 1 ML, the electron densities of excitons (iii) and (iv) are a mixture of $d_{xz}$ and $d_{y z}$ orbitals predominantly on subsurface Ti atoms below O$_{\textrm{br}}$, with some weight on surface Ti atoms below O$_{\textrm{br}}$ and \Ticus{} sites, respectively.  Exciton (iii) includes some transitions away from $\Gamma$, while the transitions of exciton (iv) are mostly away from $\Gamma$ (\textit{cf.} Table~\ref{Table3}).  For \sfrac{2}{3} ML, exciton (iii) and (iv) both have electron densities on Ti atoms throughout the substrate.  While the electron density of exciton (iii), which includes transitions away from $\Gamma$, has a mixed $d_{xz}$ and $d_{y z}$ character, the electron density of exciton (iv) is a much more complex mixture of Ti 3$d$ atomic orbitals, including a single transition from HOMO to a higher energy CB level away from $\Gamma$ (\textit{cf}.\ Table~\ref{Table3}).  

%Overall, we find that the observed
Overall, we find the brightest excitons tend to be composed of transitions away from $\Gamma$ that have electron density on both surface and subsurface Ti atoms throughout the substrate.  Excitons which are composed of a single transition at $\Gamma$ are delocalized throughout the entire system, and do not decay.  This is because the electron and hole densities are each composed of a single periodic KS orbital.  In other words, such transitions are from the entire overlayer to the substrate. It is only when multiple transitions with differing phase factors, that is, away from $\Gamma$, are included that the electron and hole may be localized and decay through destructive interference between the KS wave functions.  In this way the exciton may have the hole located on, e.g., the O 2$p_\pi$ orbital of a single catechol molecule, with the electron density decaying into the substrate.  

\begin{figure}%[!b]
%\noindent{\color{StyleColor}{\rule{\columnwidth}{1pt}}}
\includegraphics[width=\columnwidth]{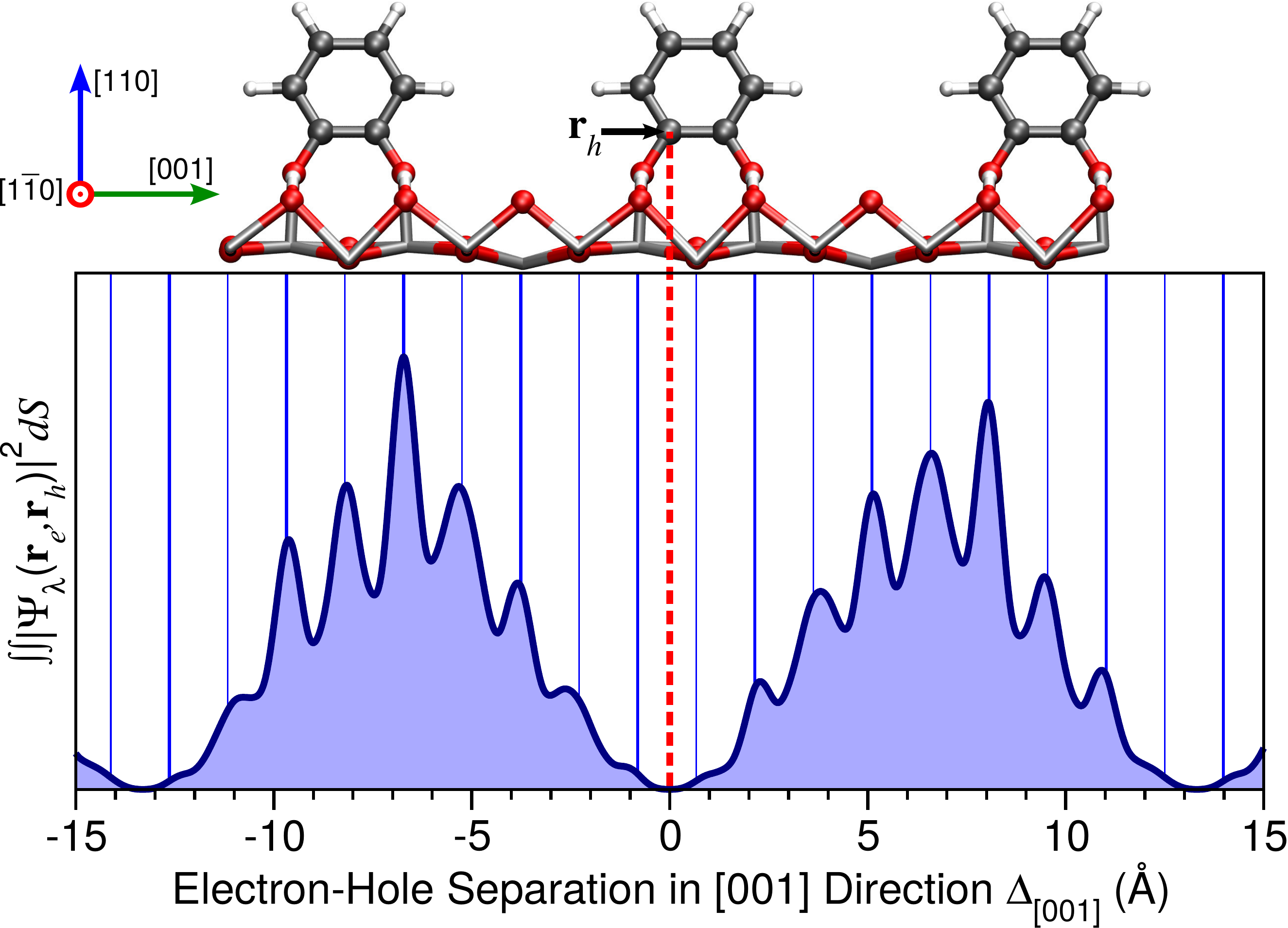}
\caption{Squared planar averaged excitonic wave function $\iint|\Psi_\lambda(\textbf{r}_e,\textbf{r}_h)|^2d S$ versus electron-hole separation in the [001] direction $\Delta_{[001]}$ in \AA, fixing the hole position (red vertical dashed line) at the maximum of the hole density, $\max_{\mathbf{r}_h}\rho_h(\mathbf{r}_h)$, for exciton (iv) of the fully dissociated \sfrac{2}{3} ML catechol on TiO$_2(110)$ shown in Figure~\ref{fgr:StructBSE}(d). The positions of Ti atoms, \Ticus{} and Ti--O$_{\textrm{br}}$, are denoted by blue vertical lines, thick and thin, respectively.}\label{fgr:Exciton3div}
\noindent{\color{StyleColor}{\rule{\columnwidth}{1pt}}}
\end{figure}

This may be important for excitons composed of transitions away from $\Gamma$, such as exciton (iv) of Figure~\ref{fgr:StructBSE}(d).  This exciton's squared planar averaged excitonic wave function in the [001] direction, $\iint  |\Psi_\lambda(\textbf{r}_e, \textbf{r}_h)|^2 d S$, is shown in Figure~\ref{fgr:Exciton3div}.  Here, the calculated Bloch wave function, with three $k$-points in the [001] direction, has been unfolded into real space.  This representation of $\Psi_\lambda$ has by construction a periodicity of 26.622~\AA{} in the [001] direction.  Increasing the number of $k$-points in the [001] direction would remove this constraint, potentially removing the nodes in the excitonic wave function at $\Delta_{[001]} \approx \pm 13.311$~\AA{}  in Figure~\ref{fgr:Exciton3div}.  We have fixed the hole position $\textbf{r}_h$ to the maximum of the hole density $\rho_h$, which is on the outer $p$ lobe of a C atom bonding to the anchoring O atom, as shown in Figure~\ref{fgr:Exciton3div}.  

We clearly see from Figure~\ref{fgr:Exciton3div}, that although exciton (iv) for the fully dissociated \sfrac{2}{3} ML catechol overlayer is very intense (\textit{cf}.~Figure~\ref{fgr:StructBSE}(d)), the electron and hole are spatially separate.  Moreover, the majority of the electron density is on Ti atoms below a catechol molecule adjacent to that on which the hole is located.  This means exciton (iv) of Figure~\ref{fgr:StructBSE}(d) is a charge transfer transition in both the [001] and [110] directions.  This suggests electron-hole separation should be facile for this bright exciton.

\section{CONCLUSIONS}\label{Sect:Conclusions}

We have performed HSE06-based BSE calculations of the optical absorption spectra and bright excitons for the most stable catechol overlayers on rutile TiO$_2$(110) as a function of coverage and dissociation. The catechol--TiO$_2$(110) interface is a type-II DSSC, with charge transfer transitions from the HOMO of catechol directly to the CB Ti 3$d$ levels of the TiO$_2$ substrate.  Our calculated spectra agree semi-quantitatively with the measured spectra for catechol on anatase TiO$_2$ nanoparticles. However, our results, combined with the lower direct electronic band gap of bulk rutile compared to anatase TiO$_2$\cite{SunH2OAnatase}, suggest that the absorption spectra of catechol on rutile TiO$_2$(110) has a better overlap with the visible region of the solar spectra than catechol on anatase TiO$_2$ nanoparticles.  Thus, the catechol--TiO$_2$(110) interface may provide a greater photovoltaic efficiency than catechol on anatase TiO$_2$ nanoparticles.  This provides motivation for future experimental studies of the optical absorption of the catechol--TiO$_2$(110) interface.

The computation of the absorption spectra, excitonic binding energies, and spatial distribution of the excitons provides valuable information for discussing how facile charge transfer and electron-hole separation is in type-II DSSCs. As the catechol overlayer is dissociated, the onset of the absorption spectrum shifts to lower energies.   The lowest-energy excitons are rather bright, with low binding energies, and consist of charge transfer transitions from the overlayer's HOMO to the substrate's CBM.  In agreement with transient absorption spectra\cite{CatecholTiO2NPTJPCB2003} we find that these excitons have different degrees of charge separation onto either surface or subsurface Ti atoms.

The relationship between intensity and charge transfer is rather complex, and one must consider the directional dependence of the absorption spectra.   Absorption along the [110] direction is related to charge transfer from the catechol overlayer to the substrate, while absorption along the [001] direction is related to charge transfer to adjacent catechol adsorption sites.  For all structures we find the strongest absorption is in the [110] direction.  

 The \sfrac{2}{3} ML catechol overlayer has a significantly stronger absorption normal to the surface plane than the 1 ML catechol overlayers.  The brightest exciton for fully dissociated \sfrac{2}{3} ML catechol on TiO$_2$(110) is a charge transfer transition with the electron and hole spatially separated in both the [110] and [001] directions.  This suggests electron conduction is most probably via a superposition of $d_{xz}$ and $d_{y z}$ orbitals aligned in the [$1\overline{1}0$] direction. 

 Overall, our results suggest the excitons of the catechol--TiO$_2$(110) interface are spread out over several unit cells.  This is based on their weak binding energies, the lower energy excitons being predominantly composed of transitions at $\Gamma$, and the weak dispersion of rutile TiO$_2$'s band structure.  Such excitons would be difficult to describe with small nanoparticle models for the substrate.  This work motivates future calculations with denser $k$-point meshes and thicker substrate models to determine whether the bright excitons of the catechol--TiO$_2$ interface have Frenkel or Wannier character.

\section*{ASSOCIATED CONTENT}
\subsubsection*{\includegraphics[height=8pt]{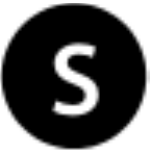} Supporting Information}
\noindent
%Comparison of $\rho(\textbf{r},\varepsilon_F+U)$ and $\int_{\varepsilon_F}^{\varepsilon_F+U}\rho(\textbf{r},\varepsilon)d\varepsilon$ linescans.  
The Supporting Information is available free of charge on the \href{http://pubs.acs.org}{ACS Publications website} at DOI: \href{http://pubs.acs.org/doi/abs/10.1021/acs.jctc.6b00217}{10.1021/acs.jctc.6b00217}.

\hangindent=0.33cm Comparison of IP and BSE optical absorption spectra, total energies, and atomic coordinates of all structures (\href{http://pubs.acs.org/doi/suppl/10.1021/acs.jctc.6b00217/suppl_file/ct6b00217_si_001.pdf}{PDF})

\section*{%%\large$\blacksquare$\normalsize\ 
AUTHOR INFORMATION}
\subsubsection*{Corresponding Author}
\noindent *E-mail: \href{mailto:duncan.mowbray@gmail.com}{duncan.mowbray@gmail.com}.\\
\noindent *E-mail: \href{mailto:annapaola.migani@icn2.cat}{annapaola.migani@icn2.cat}.
\subsubsection*{Notes} 
\noindent The authors declare no competing financial interest.
\section*{%%large$\blacksquare$\normalsize\ 
ACKNOWLEDGMENTS} 

We acknowledge financial support from Spanish Grants (FIS2012-37549-C05-02, FIS2013-46159-C3-1-P, RYC-2011-09582, JCI-2010-08156); Generalitat de Catalunya (2014SGR301, XRQTC); Grupos Consolidados UPV/EHU del Gobierno Vasco (IT-578-13).
\bibliography{bibliography}
%\tocentry{\begin{center}\includegraphics[width=5.1cm]{ExcitingDyesTOC}\end{center}}

\end{document}